\documentclass[conference]{IEEEtran}
\usepackage[utf8]{inputenc}
\usepackage[T1]{fontenc} 

\usepackage{booktabs}
\usepackage{tabularx}

\usepackage{cite}

\ifCLASSINFOpdf
  \usepackage[pdftex]{graphicx}
  \graphicspath{{figures/}}
  \DeclareGraphicsExtensions{.pdf,.jpeg,.png}
\else
  \usepackage[dvips]{graphicx}
  \graphicspath{{figures/}}
  \DeclareGraphicsExtensions{.eps}
\fi

\usepackage{amsmath}

\usepackage{siunitx}
\ifCLASSOPTIONcompsoc
  \usepackage[caption=false,font=normalsize,labelfont=sf,textfont=sf]{subfig}
\else
  \usepackage[caption=false,font=footnotesize]{subfig}
\fi

\usepackage{stfloats}

\usepackage{hyperref}

\newcommand\figref{\figurename~\ref}
\newcommand\secref{Section~\ref}

\usepackage{tikz}
\usetikzlibrary{calc, positioning}
\newif\ifpreprint

\def\conferenceyear{2019}                                       
\def\conferencemonth{September}                                      
\def\conferencetitle{IEEE Vehicular Technology Conference}      
\def\conferencetitleabrv{VTC-Fall}                            
\def\conferencelocation{Honolulu, Hawaii}                        

\def\conferencenotice{
	Accepted for presentation in: \conferencetitle~(\conferencetitleabrv), \conferencelocation, \conferencemonth~\conferenceyear.}
\def\copyrightnotice{
	\copyright~\conferenceyear~IEEE. Personal use of this material is permitted. Permission from IEEE must be obtained for all other uses, including reprinting/republishing this material for advertising or promotional purposes, collecting new collected works for resale or redistribution to servers or lists, or reuse of any copyrighted component of this work in other works.}

\preprinttrue

\begin{document}
%
\title{Performance Analysis of C-V2X Mode 4 Communication Introducing an Open-Source C-V2X Simulator}


%
\author{\IEEEauthorblockN{
Fabian Eckermann, Moritz Kahlert and Christian Wietfeld}
\IEEEauthorblockA{
TU Dortmund University, Communication Networks Institute (CNI)\\
Otto-Hahn-Str. 6, 44227 Dortmund, Germany\\
E-Mail:\{fabian.eckermann, moritz.kahlert, christian.wietfeld\}@tu-dortmund.de}}

\maketitle

\ifpreprint
    \begin{tikzpicture}[remember picture, overlay]
    	\node[below=5mm of current page.north, text width=20cm,font=\sffamily\footnotesize,align=center] {\conferencenotice};
    	\node[above=5mm of current page.south, text width=15cm,font=\sffamily\footnotesize] {\copyrightnotice};
    \end{tikzpicture}%
\fi

\begin{abstract}
Autonomous vehicles, on the ground and in the air, are the next big evolution in human mobility. While autonomous driving in highway scenarios is already possible using only the vehicles sensors, the complex scenarios of big cities with all its different traffic participants is still a vision. Cellular Vehicle-to-Everything (C-V2X) communication is a necessary enabler of this vision and and an emerging field of interest in today's research. However, to the best of our knowledge open source simulators essential for open research do not exist yet. In this work we present our open source C-V2X mode 4 simulator based on the discrete-event network simulator ns-3. To analyze the performance of C-V2X mode 4 using our simulator, we created a worst case scenario and the 3GPP reference Manhattan grid scenario using the microscopic traffic simulator SUMO. We also added the WINNER+ B1 channel model to ns-3, as this is also used by 3GPP. Our results show, that C-V2X is scalable to 250 vehicles within a worst case scenario on a playground of 100~m x 100~m, with respect to the LTE rel. 14 V2X requirements. For the more realistic Manhattan grid scenario, the performance is better, as to be expected. We also analyzed the Packet Inter-Reception time with an outcome of max. 100~ms for more than 99~\% of all transmissions. In addition, we investigated the impact of the Resource Reservation Period and the Resource Reselection Probability on the system's Packet Reception Ratio.
\end{abstract}


\IEEEpeerreviewmaketitle

\section{Introduction} \label{intro}

\begin{figure}[t]
    \centering
    \includegraphics{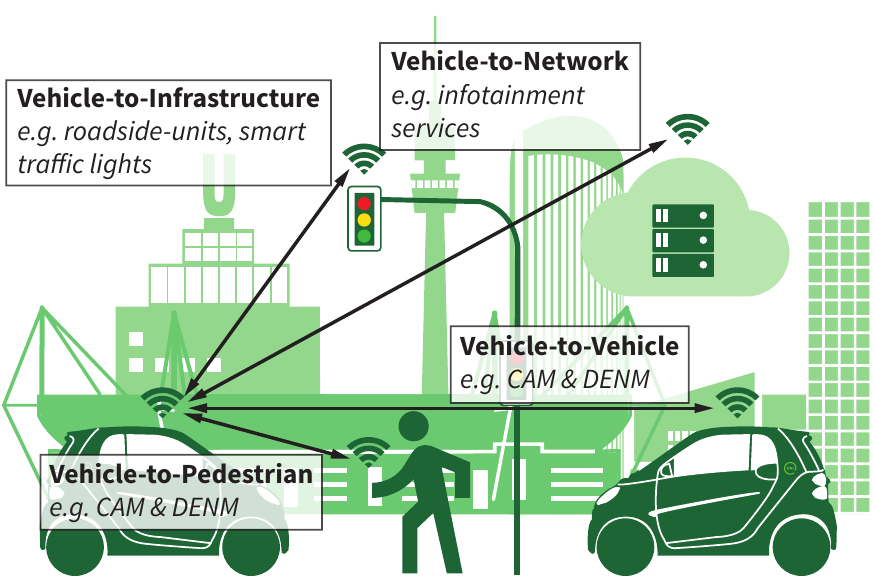}
    \caption{C-V2X use cases: Vehicle-to-Vehicle (V2V), Vehicle-to-Pedestrian (V2P), Vehicle-to-Infrastructure (V2I) and Vehicle-to-Network (V2N).)}
    \label{fig:scenario}
\end{figure}

Autonomous vehicles are a necessity to reduce the worldwide traffic load. A UK report from 2012 found out that the average car is parked for 96~\% of the time \cite{bates2012spaced}. In contrast to being parked, an autonomous vehicle would be available to other family members, even those who cannot drive on their own e.g. children or disabled people reducing the need for multiple cars per family.\\
Vehicular communication is a very important enabler to autonomous vehicles with two competing technologies. A WiFi-based standard for vehicular communication, IEEE 802.11p, was introduced in 2010, the 3GPP Cellular Vehicle-to-Everything (C-V2X) standard was released in 2017 (rel. 14). The 3GPP specifies four types of C-V2X applications \cite{3gpp.22.185}: Vehicle-to-Vehicle (V2V), Vehicle-to-Pedestrian (V2P), Vehicle-to-Infrastructure (V2I) and Vehicle-to-Network (V2N) as shown by \figref{fig:scenario}.

While IEEE 802.11p has been comprehensively studied analytically, in simulations and field-testing by academia and industry for example in \cite{4526014, 4350110, 4657278, 5720204, 1023879, 4015707}, the newer C-V2X standard has not. In \cite{2017AnAO}, a study is presented that compares C-V2X communication with IEEE 802.11p stating that using C-V2X communication can avoid a higher number of fatalities and serious injuries due to its superior reliability.\\
For C-V2X two modes, extending mode 1 and mode 2 that were defined in LTE Device-to-Device communication, are defined. Mode 3, only available under network coverage, where a base station schedules the sidelink resources and mode 4, that is unsupervised (see \secref{C-V2X}). While C-V2X simulators have been developed and introduced in different studies, to the best of the authors knowledge, none of them is available open source.\\
An open source analytical model for C-V2X mode 4 is introduced in \cite{8581518}. Analytical models for the reliability as a function of the distance between the communication nodes and for different transmission errors that can be encountered in C-V2X mode 4 are presented and validated for a multitude of transmission parameters and traffic densities. A validation of the analytical models by a comparison to simulation results is presented in \cite{8080373} and \cite{8108463}. In \cite{8080373} the performance of C-V2X mode 4 is analyzed for a highway scenario defined by 3GPP \cite{3gpp.36.885}. A detailed performance analysis is presented and a modification to the scheduling is proposed. In \cite{8108463} the performance of C-V2X mode 4 is evaluated for an urban scenario under realistic traffic conditions. Furthermore, it is shown, that the sensing-based Semi-Persistent Scheduling (SPS) scheme improves the performance compared to a random resource selection. The introduced simulator focuses on the evaluation of the sensing-based Semi-Persistent Scheduling of C-V2X mode 4.\\
Another C-V2X simulator is presented in \cite{180410788}. The authors investigate the influence of resource pool parameters transmission configurations for C-V2X mode 4 on the reliability of the communication.\\
In \cite{8628416}, a performance analysis of the resource allocation of the C-V2X multiple access mechanism is presented.  The authors use a simulator based on the ns-3 LTE module to evaluate the system level performance. They conclude, that optimum performance can be achieved if unified system configurations for C-V2X are mandated, as it is for IEEE 802.11p.\\
A comparison of the aforementioned work among each other and to the results of our work is difficult, as  different performance metrics, different scenarios or different simulation configurations are used and the reasons for varying results can not be identified due to the lack of insight in the simulator implementations.
To allow for open research our simulator is published open source\footnote{\url{https://github.com/FabianEckermann/ns-3\_c-v2x}} to be validated, used and improved by other researchers. A first usage of the simulator is presented in \cite{sliwa2019lightweight}. A lightweight  framework for integrated simulation of aerial and ground-based vehicular networks is shown and the performance of LTE and C-V2X for connecting the ground- and air-based vehicles is compared.



In the remainder of this paper, we build upon this motivation and give a short overview over the principles of C-V2X communication and our performance criteria (\secref{C-V2X}). In \secref{system}, the system design and the simulation parameters of our C-V2X simulator and the implemented channel model used for our simulations are explained. \secref{experiment} contains the simulation results of different case studies and we finally conclude the paper in \secref{conclusion}.

\section{C-V2X Communication} \label{C-V2X}

In this section a short overview of the relevant C-V2X background is given and the performance criteria used in this paper are described. 

\subsection{Principles of C-V2X Communication Mode 4}

For C-V2X mode 4, sensing-based Semi-Persistent Scheduling (SPS) is introduced as distributed scheduling protocol by 3GPP, to autonomously select radio resources. The protocol takes advantage of the periodic and predictable nature of V2X communication services.
With sensing-based SPS V-UEs (Vehicular User Equipment) reserve subchannels in the frequency domain for a random number of consecutive periodic transmissions. The number of reserved subchannels per subframe depends on the size of data to be transmitted.
\begin{figure}
    \centering
    \includegraphics{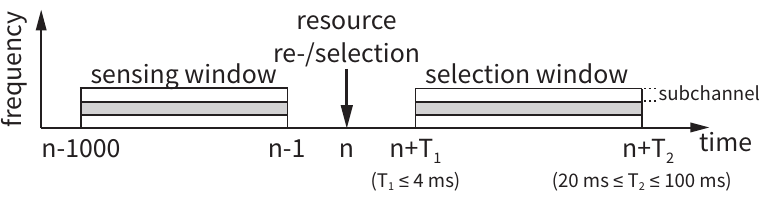}
    \caption{Sensing-based Semi-Persistent Scheduling}
    \label{fig:sps}
\end{figure}
A timeline of the resource allocation mechanism is illustrated in \figref{fig:sps}. A V-UEs resource selection (or reselection) at time $n$ is based on the received data from other vehicles within the past 1000~ms (sensing window). Resources for transmission are then picked within the selection window from a list of resources that are estimated to not be occupied by the periodic transmissions of the other vehicles. The lower bound of the selection window ($T_1$) depends on the V-UEs configuration while the upper bound is defined by the maximum Packet Inter-Reception time allowed for the type of transmission. If all resources are occupied the V-UE will transmit on occupied resources with low received power, as the importance of vehicular communication services is highly related to its direct surrounding.\\
After a resource is selected, the V-UE will transmit periodic messages. The transmission interval is defined by the Resource Reservation Period. After using the resource for 5 to 15 transmissions a resource reselection based on the Resource Reselection Probability (in the interval of [0.2, 1]) is triggered.

\subsection{Performance Criteria}
For the validation of the simulation results the Packet Reception Ratio (PRR) and Packet Inter-Reception (PIR) as specified in \cite{3gpp.36.885} are used.
\begin{itemize}
    \item The PRR is calculated by $X/Y$, where $Y$ is the number of V-UEs that are located in the baseline distance $(a, b)$ from the transmitter, and $X$ is the number of V-UEs with successful packet reception among $Y$.
    \item The PIR describes the time between two successful receptions of two different successive packets transmitted from node A to node B.
\end{itemize}
The transmission latency is not considered within this work. The propagation delay for the specified C-V2X effective distances of 320~m (freeway/motorway) \cite{3gpp.22.885} is $\approx1~\mu$s. Further delays due to processing times are not included in the simulations. 

\section{System Model and Scenarios} \label{system}
%

\begin{figure}
    \centering
    \includegraphics{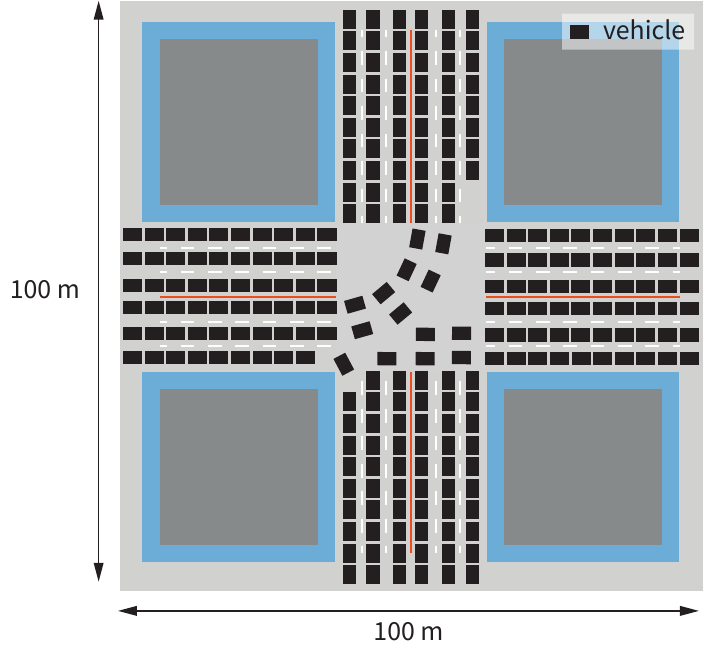}
    \caption{Static intersection scenario with 250 V-UEs.}
    \label{fig:static}
\end{figure}

\begin{figure}
    \centering
    \includegraphics{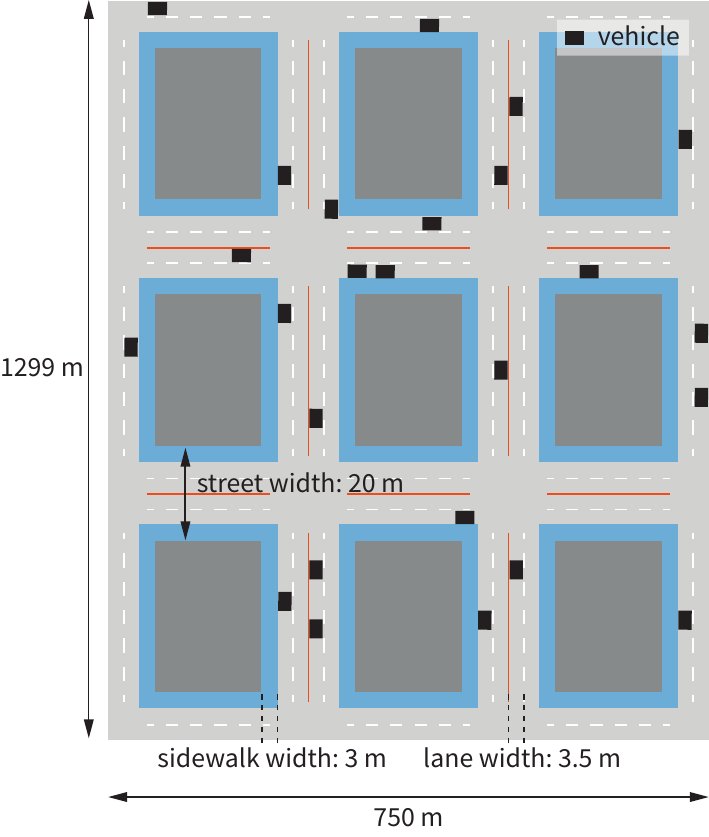}
    \caption{Manhattan grid for the urban simulation use cases as used by 3GPP.}
    \label{fig:manhattan}
\end{figure}

Our C-V2X simulator is based on the LTE Device-to-Device communication simulation model for the ns-3 network simulator introduced and validated in \cite{Rouil2017}. It includes mode 1 and mode 2 direct communication and in this work C-V2X Mode 4 has been added to the ns-3 simulator. We implemented the adjacent and non-adjacent Resource Block (RB) assignment schemes and the sensing-based SPS. Vehicular traffic is simulated by the microscopic traffic simulator SUMO \cite{Sumo2018}. For the mobility simulation of the vehicular traffic two scenarios are used. A static worst case scenario where all vehicles are placed on a 100~m x 100~m intersection such that all transmitted packets will be received by all vehicles, (see \figref{fig:static}) and an urban Manhattan grid scenario as used by 3GPP \cite{3gpp.36.885}, shown in \figref{fig:manhattan}.
%
Unless otherwise noted, the simulation parameters are set as listed by TABLE I.
In addition, the WINNER+ B1 has been added to the ns-3 simulator in order to align with the 3GPP studies and to achieve comparable results.
\subsection{WINNER+ B1 Channel Model}
 For the studies on LTE-based V2X services, the WINNER+ B1 channel model
 is used by the 3GPP \cite{3gpp.36.885}. The WINNER+ channel models are an update of the WINNER II channel models presented in \cite{winner2}.
The B1 channel model for the 5.9~GHz band is calculated as follows:\\
\begin{itemize}
    \item LOS:
    for $30~m < d < d'_{BP}$:
    \begin{equation}
        \noindent\hspace*{4pt}PL(dB) = 22.7 \cdot log_{10}(d) + 27.0 + 20.0 \cdot log_{10}(f_c)
    \end{equation}
    for $d'_{BP} < d < 5~km$:
    \begin{equation}
        \begin{split}
        PL(dB) = & 40.0 \cdot log_{10}(d) + 9 - 16.2 \cdot log_{10}(h_{BS})\\
        & - 16.2 \cdot log_{10}(h_{MS} + 3.8 \cdot log_{10}(f_c)
        \end{split}
    \end{equation}
    \item NLOS:
    \begin{equation}
        \begin{split}
        PL(dB) = (44.9 - 6.55 \cdot log_{10}(h_{BS})) \cdot log_{10}(d)\\
        + 5.83 \cdot log_{10}(h_{BS}) + 15.38 + 23 \cdot log_{10}(f_c)
        \end{split}
    \end{equation}
\end{itemize}
The effective breakpoint distance $d'_{BP}$ is calculated by:
\begin{equation}
    \begin{split}
    d'_{BP} = 4 \cdot h'_{BS} \cdot h'_{MS} \cdot f_c/c
    \end{split}
\end{equation}
where the effective antenna heights $h'_{BS} = h_{BS} - 1~m$ and $h'_{MS} = h_{MS} - 1~m$ and $c = 3\cdot10^8~m/s$.\\
As defined by \cite{3gpp.36.885} the antenna height for V2V communication should be set to $h_{BS} = h_{MS} = 1.5~m$.
\begin{table}[h!]
\caption{Simulation parameters}
\label{tab:config}
\centering
\setlength{\tabcolsep}{10pt} 
\renewcommand{\arraystretch}{1}
    \begin{tabular}{ll}
        \toprule
        \textbf{\textsc{general parameters}}      &\\
        \toprule
        number of V-UEs                           & 0...250\\
        channel model urban                       & WINNER+ B1\\
        baseline distance (urban)                 & 150~m\\
        ns-3 version                              & 3.28\\
        SUMO version                              & 0.32.0\\
        simulation time                           & 30~s\\
        \toprule
        \textbf{\textsc{V-UE parameters}}         &\\
        \toprule
        message size                              & 190~bytes\\
        transmission power                        & 23~dBm\\
        antenna height                            & 1.5~m\\
        resource reservation period               & 100~ms\\
        $T_1$, $T_2$                              & 4~ms, 100~ms\\
        resource reselection probability          & 50~\%\\
        modulation and coding scheme              & 20\\
        \toprule
        \textbf{\textsc{resource pool parameters}}&\\
        \toprule
        channel bandwidth                         & 10, 20~MHz\\
        RBs per subchannel                        & 10\\
        number of subchannels                     & 5, 10\\
        subframe bitmap                           & 0xFFFFF\\
        subchannel scheme                         & adjacent\\
        Lowest RB subchannel index                & 0\\
        \bottomrule
    \end{tabular}
\end{table}%

\section{Simulation Results} \label{experiment}
\begin{figure}
    \centering
    \includegraphics{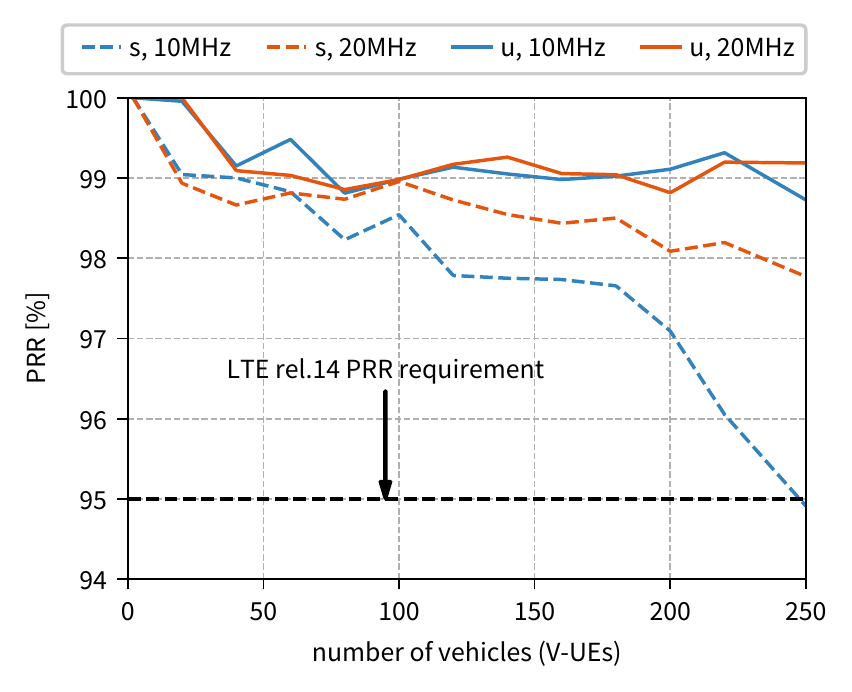}
    \caption{Packet Reception Ratio for an increasing number of V-UEs and cellular bandwidths of 10~MHz and 20~MHz on the static playground (s, 100~m x 100~m) and the urban Manhattan playground (u, 750~m x 1299~m).}
    \label{fig:prr_static_urban}
\end{figure}
\begin{figure}
    \centering
    \subfloat[][Static intersection scenario (100~m x 100~m). \label{fig:pir_static}]{%
        \includegraphics{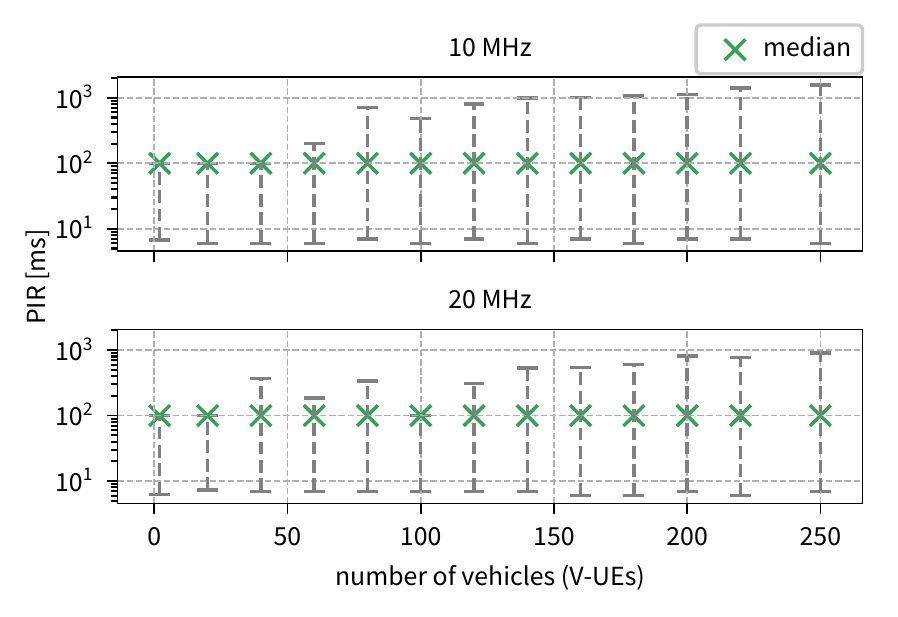}
    }
    \hfill
    \subfloat[][3GPP Manhattan grid reference scenario (750~m x 1299~m). \label{fig:pir_urban}]{%
        \includegraphics{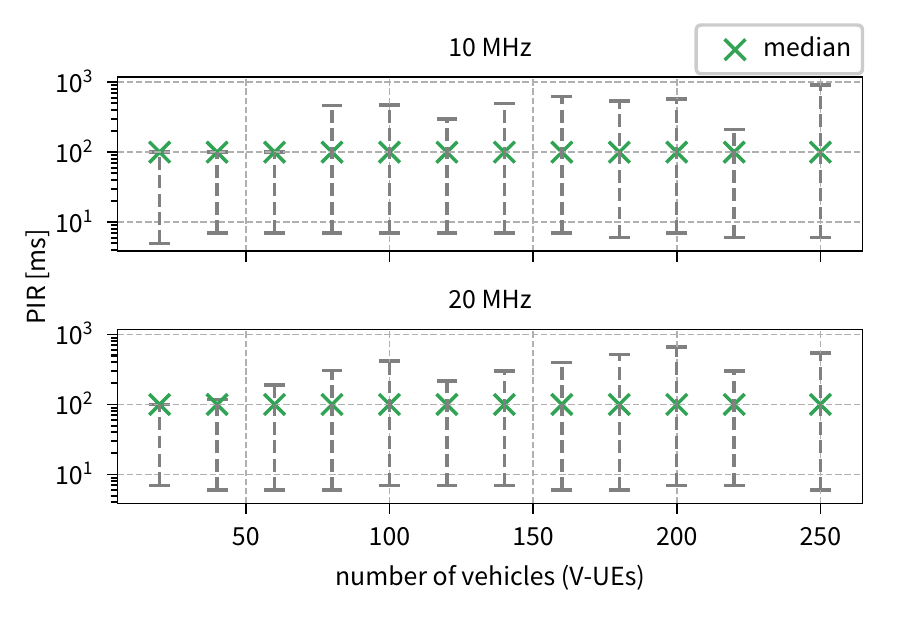}
    }
    \caption{Packet Inter-Reception for an increasing number of V-UEs and cellular bandwidths of 10~MHz and 20~MHz. Note that the whiskers mark the 0.1~\% and 99.9~\% quantiles. The maximum PIR of the 99~\% quantile is 100~ms for all these simulation runs.}
    \label{fig:pir}
\end{figure}
\begin{figure}
    \centering
    \includegraphics{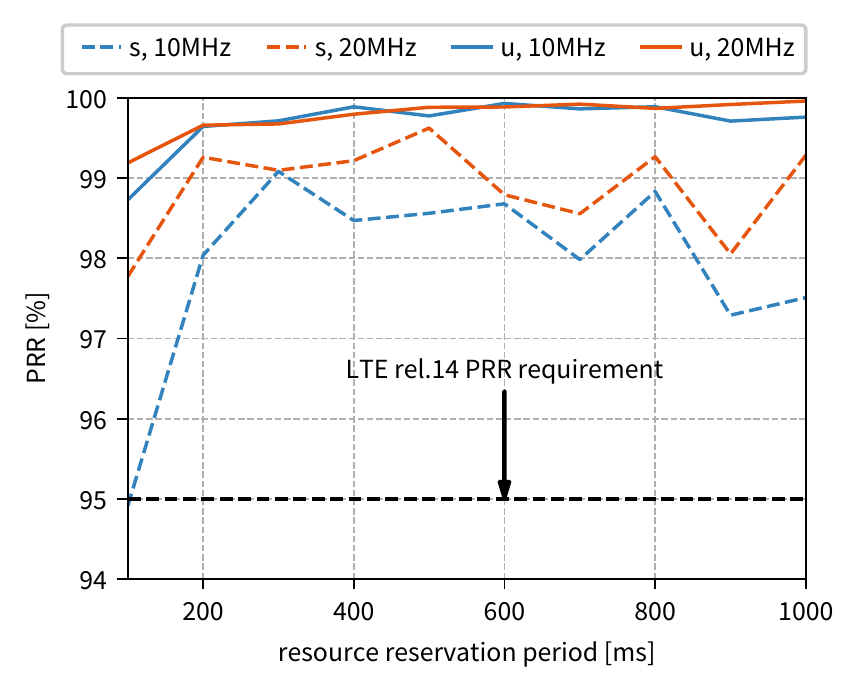}
    \caption{Packet Reception Ratio (PRR) for increasing Resource Reservation Period and cellular bandwidths of 10~MHz and 20~MHz on the static playground (s, 100~m x 100~m) and the urban Manhattan playground (u, 750~m x 1299~m) for 250 vehicular UEs (V-UEs).}
    \label{fig:prr_pRsvp}
\end{figure}

V2X communication must be scalable to guarantee road safety even for inner city scenarios with massive traffic volume. \figref{fig:prr_static_urban} shows the outcome of a scalability analysis for a static scenario and the 3GPP reference Manhattan grid. In the static scenario with 10~MHz cellular bandwidth the LTE Rel. 14 requirements \cite{3gpp.22.885} can be fulfilled for up to ~250 vehicles. This is a worst case scenario, as all of the up to 250 vehicles are always in communication range of each other, so the communication channel is loaded. If the cellular bandwidth is doubled even for 250 vehicles a packet reception ratio of ca. 98~\% is achieved. For a more realistic, dynamic scenario on a bigger playground the packet reception ratio is even for 250 vehicles and cellular bandwidths of 10~MHz and 20~MHz at around 99~\%, as only several vehicles are within the communication range of other vehicles.\\
Another important performance indicator used by 3GPP is the Packet Inter-Reception (PIR). This is analyzed for an increasing number of vehicles in \figref{fig:pir_static} for the static worst case scenario and in \figref{fig:pir_urban} for the dynamic Manhattan grid scenario. The PIR depends on the Resource Reservation Period that is set to 100~ms. Lower PIR times occur if new resources are selected by the vehicle. The resulting PIR values of new resources are selected are within the range of the selection window $T_1 - T_2$ (4~ms - 100~ms). PIR times exceeding 100~ms can occur due to half-duplex or collision errors. As the PIR of the 99~\% quantile is 100~ms for all analyzed simulation runs, the 99.9~\% quantile is shown. In the static scenario using a cellular bandwidth of 10~MHz, the PIR for the 99.9~\% quantile exceeds 1~s if more than 200 vehicles are simulated. If the cellular bandwidth is doubled, the PIR of the 99.9~\% quantile is lowered. For the dynamic Manhattan scenario, the PIR is also improved for the same reasons as for the improved scalability. It is shown, that a reliable PIR of max. 100~ms can be achieved for 99~\% of all transmitted messages. 
In \figref{fig:prr_pRsvp} the impact of the resource reservation period on the reliability is investigated. If a resource reservation period of 100~ms is set for C-V2X communication, the PRR for a communication scenario with 10~MHz cellular bandwidth and 250 vehicles drops below 95~\%. If the resource reservation period is doubled to 200~ms, the PRR increases to about 98~\%. This is expected, as the doubling of the resource reservation period corresponds to a scenario with half the traffic load that is 125 vehicles where, as shown by \figref{fig:prr_static_urban} the PRR is about 98~\% as well. For the more realistic urban Manhattan scenario the PRR increases from about 99~\% to almost 100~\%.%
\begin{figure}
    \centering
    \includegraphics{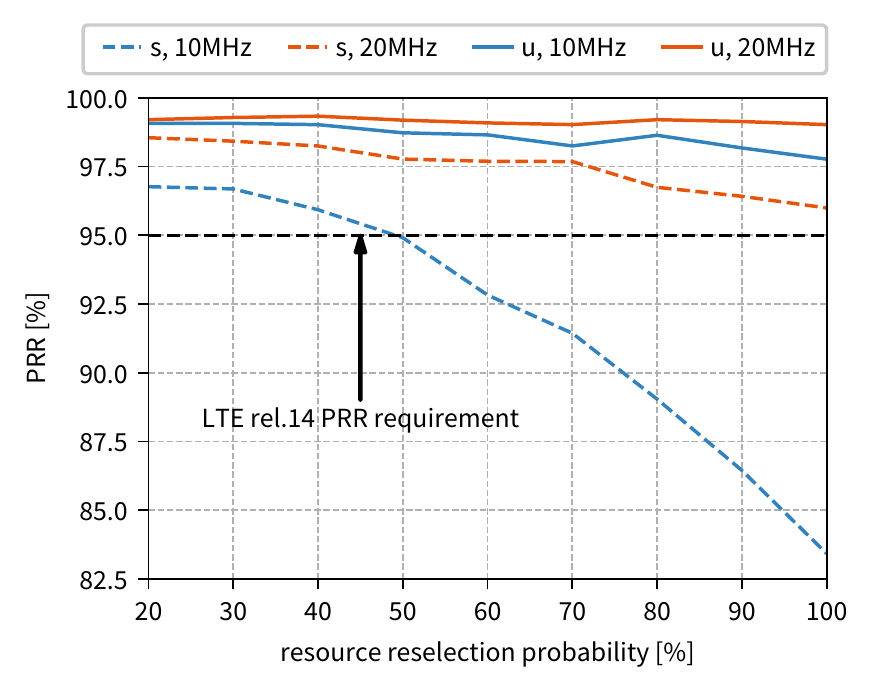}
    \caption{Packet Reception Ratio (PRR) for increasing Resource Reselection Probability and cellular bandwidths of 10~MHz and 20~MHz on the static playground (s, 100~m x 100~m) and the urban Manhattan playground (u, 750~m x 1299~m) for 250 vehicular UEs (V-UEs).}
    \label{fig:prr_rrp}
\end{figure}
Last, we analyzed the Packet Reception Ratio for an increasing Resource Reselection Probability (\figref{fig:prr_rrp}). If the Resource Reselection Probability is high, the sensing-based SPS does not perform well, as new resources are selected more often so the resource selection process becomes less predictable. This leads to a degradation of the communication reliability, as the results depict. A lower Resource Reselection Probability on the other hand leads to a more predictable channel usage, but also increases the probability of burst-like errors and leads to vehicles that might never communicate with each other if they picked the same resources once.

\section{Conclusion} \label{conclusion}

In this paper we introduced and evaluated an open source C-V2X mode 4 simulator implemented in ns-3. We analyzed the performance of our simulator using the performance criteria of the Packet reception Ratio and the Packet Inter-Reception. Our simulation results show that C-V2X mode 4 is highly scalable even for worst case scenarios. Within more realistic scenarios the performance is even better and above the 3GPP rel. 14 V2X requirements. We also analyzed the impact of the Resource Reservation Period and the Resource Reselection Probability on the system performance. Further work is underway regarding a further implementation of C-V2X mode 3 and simulations using real world scenarios. We look forward to see researchers testing, using and improving this simulator to gather further knowledge and advance.
\newpage

\section*{Acknowledgment}
\scriptsize{
The work on this paper has  been  partially funded by the federal state of Northrhine-Westphalia and the “European Regional Development Fund” (EFRE) 2014-2020 in the course of the InVerSiV project under grant number EFRE-0800422 and by Deutsche Forschungsgemeinschaft (DFG) within the Collaborative Research Center SFB 876 project B4.
}



\bibliographystyle{IEEEtran}


\end{document}